\RequirePackage[2020-02-02]{latexrelease}
\documentclass[twocolumn,prl,amsmath,amssymb,showpacs,superscriptaddress,floatfix]{revtex4}
\usepackage{graphicx}
\usepackage{bm}
\usepackage{lineno,hyperref}
\usepackage{epstopdf}
\usepackage{epsf}
\usepackage{xcolor}

\begin{document}
\title{Molecular simulation of bulk and confined (1,1,1,3,3-pentafluorobutane)}

\author{Yu. D. Fomin \footnote{Corresponding author: fomin314@mail.ru}}
\affiliation{Vereshchagin Institute of High Pressure Physics,
Russian Academy of Sciences, Kaluzhskoe shosse, 14, Troitsk,
Moscow, 108840, Russia }
\affiliation{The Sophia Kovalevskaya North-West Center for Mathematical Research Center, Immanuel Kant Baltic Federal University, 236016 Kaliningrad, Russia
}

\author{E. N. Tsiok}
\affiliation{Vereshchagin Institute of High Pressure Physics,
Russian Academy of Sciences, Kaluzhskoe shosse, 14, Troitsk,
Moscow, 108840, Russia}

\author{V. N. Ryzhov}
\affiliation{Vereshchagin Institute of High Pressure Physics,
Russian Academy of Sciences, Kaluzhskoe shosse, 14, Troitsk,
Moscow, 108840, Russia}
\date{\today}

\begin{abstract}

Here we present a computational study of the thermodynamic and
structural properties of bulk and confined
(1,1,1,3,3-pentafuorobutane) with different lengths of the carbon
backbone. The DREIDING force field model has been used in the
method of molecular dynamics. In order to study the effect of
confinement we have placed (1,1,1,3,3-pentauorobutane) molecules
between two graphene walls. In order to study the influence of
pore loading on system behavior we have simulated systems of the
same size, but with a different number of
(1,1,1,3,3-pentauorobutane) molecules, from 200 to 2000. The
equations of state at $T = 300$ K in a wide range of densities for
all considered systems had a single peculiarity that is attributed
to gas-liquid transition. From the two-dimensional radial
distribution functions, density profile and angular distribution
we have observed the systems split into layers with amorphization
rather than crystallization in them.

\end{abstract}


\maketitle

\section{Introduction}

Liquids are involved in numerous natural and technological
processes. It makes the understanding of their behavior of great
interest. At the same time, many technologically important liquids
have a rather complex molecular structure. Moreover, they can form
homologous series, where the $N+1$-th term is made by adding of a
monomer to the $N$-th one. The most obvious example is the set of
hydrocarbons: $CH_4$, $CH_3-CH_3$, $CH_3-CH_2-CH_3$, etc. It is
well known that the properties of hydrocarbons are strongly
dependent on the chain length.

Hydrocarbons are the simplest organic compounds, which are used as
a fuel and raw materials for the chemical industry. It makes them
the most studied homologous seria. The properties of hydrocarbons
are well documented in numerous handbooks and databases (see, for
instance, \cite{hydrocarb}).

A more complex case is related to investigation of confined,
rather than bulk liquids \cite{rice}. In this case the properties
of the system are determined not only by the liquid itself, but
also by the shape of the confined media and interaction between
the liquid and confining walls. However, there are some effects
which are common for any kind of confinement. In particular,
confinement leads to modulation of the average density of the
system, i.e. while the local density of a bulk liquid is uniform,
in the case of confinement the interplay of dense and dilute
regions appears inside the pore \cite{rice}. Confinement also
changes the melting point of a substance. The melting temperature
of confined liquid can be either above, or below that of bulk
liquid. Moreover, the melting temperature depends on the width of
the pore and it can demonstrate non-monotonous behavior depending
on the size of the pore \cite{vish}.  Confinement may also induce
formation of crystalline structures which are not possible in a
bulk state, as it was shown in molecular simulation of water in
nanotubes and silicon in a slit pore \cite{w-nano,rinp}. One can
conclude that the behavior of confined liquids involves numerous
different phenomena, which make it an endless topic of research.
However, even in the case of hydrocarbons in confined space there
is still lack of data. Moreover, most data are obtained for small
hydrocarbons, like methane \cite{ch4}, butane to octane
\cite{butane}, pentane \cite{pisarev}, etc.

Another class of molecules which can form chains of different
length is fluorocarbons of the type $CH_3-(CF_2-CH_2)_n-CF_3$. The
simplest molecule of this type is (1,1,1,3,3-pentafluorobutane)
(PFB) $CH_3-CF_2-CH_2-CF_3$. PFB is used as a refrigerant liquid
and as a foaming liquid. However, the polymer of the composition
$(CH_2-CF_2)_n$ (PVDF) is a widely used fluoropolymer. Although
there are plenty of works on fluoropolymers \cite{handbook}, there
are still a lot of open questions. Moreover, introduction of
fluorine makes the set of homologous substances of
fluoro-hydrocarbons richer than that of hydrocarbons.

In Ref. \cite{pfb-4} the density of saturated vapor and liquid of
PFB and its thermal conductivity are measured in the temperature
interval from $T=289.15$ to $413.15$ K. The critical point is
estimated from the obtained data. The boiling curve of PFB in a
wider range of temperatures is reported in Ref. \cite{pfb-2},
while in Ref. \cite{pfb-1} the thermophysical properties of a
mixture of PFB with Perfluoropolyether were studied. However,
there is still lack of experimental data on PFB, which is to a
certain degree compensated by computational works.

In Ref. \cite{pvdf-ff-0} a force field (FF) for PVDF was
constructed. The authors used quantum chemical calculations of
small molecules like PFB in order to fit the force field. The
obtained parameters were used for molecular dynamics calculations
of thermodynamic properties of small oligomers of PVDF.

In Ref. \cite{pvdf-ff} the FF for PVDF was improved. The authors
managed to accurately reproduce the equation of state and the
boiling curve of PFB in good agreement with experimental data.

There are also a lot of computational works on the properties of
PVDF with different FFs for the case of long molecules (to name a
few, see ). However, one can see that all works deal either with
small molecules like PFB (see Ref. \cite{pvdf-ff-0} for other
small molecules), or very long molecules (for instance, 100
monomers of $CF_2-CH_2$). To the best of our knowledge there are
no works which discuss changes of the properties of systems with
the composition $CH_3-(CF_2-CH_2)_n-CF_3$ in connection with the
length of the carbon backbone.

The goal of the present work is to perform a computational study
of bulk and confined PFB within the framework of a computationally
cheap FF. We consider this work as the first step in a sequence of
works on investigation of the properties of systems of the
composition $CH_3-(CF_2-CH_2)_n-CF_3$ with different lengths of
the carbon backbone.

\section{System and Methods}

In the present paper we investigate the properties of bulk and
confined PFB by means of molecular dynamics simulation. The
DREIDING force field \cite{dreiding} is used as an interaction
model. The details of the interaction are given in the
Supplementary Materials. Figure \ref{pvdf-mol} shows a molecule of
PFB. Each carbon atom in the molecule is treated as a separate
type ($C_1$ to $C_4$). Fluorines (hydrogen) atoms in the $CF_3$
($CH_3$) and $CF_2$ ($CH_2$) groups have different charges. As a
result, 8 atom types are used to describe the interaction.

\begin{figure}

\includegraphics[width=8cm, height=6cm]{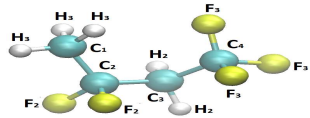}%

\caption{\label{pvdf-mol} A view of a molecule of
(1,1,1,3,3-pentafluorobutane). The symbols next to the atoms show
the types of these atoms.}
\end{figure}

In the case of simulation of bulk PFB a system of 1000 molecules
in a cubic box with periodic boundary conditions is used. The
system is simulated for $5 \cdot 10^7$ steps with a time step of
$dt=0.1$ fs for equilibration and more $5 \cdot 10^7$ steps for
calculation of the properties of the system. The bonds involving
hydrogen atoms are constrained by the SHAKE algorithm
\cite{shake}.

In order to characterize the thermodynamic properties of the
system we calculate the equation of state and compare it with
literature data. The structure of the system is monitored via
partial radial distribution functions.

To study the effect of confinement we place PFB molecules between
two graphene walls (the distance between the walls is $30 \AA$).
Each wall consists of $4200$ carbon atoms and has the sizes
$L_x=103.32$ $\AA$ and $L_y=106.54$ $\AA$. The box was periodic in
the x and y directions, but not z. The interaction parameters
between the PFB molecules and the walls are given in the
Supplementary materials.

For to see the influence of pore loading on the behavior of the
system we simulated systems of the same size, but with a different
number of PFB molecules. The smallest system consisted of 200 PFB
molecules, while the largest one consisted of 2000 molecules.
Following Ref. \cite{pisarev} we performed the Voronoi
construction for the particles of PFB and used the sum of the
volumes of each particle to calculate the density of the liquid.

In an effort to evaluate the influence of the walls on the
structure of the system we calculated the density profile of all
species. A density profile is defined as a number of particles of
a particular type inside the slab between $z$ and $z+dz$. We also
studied the angular distribution of the molecules within these
slabs. We define angle $\theta$ as the angle between the vector
connecting the $C_1$ and $C_4$ atoms and the z axis.

We observe that the system splits into layers. In order to
characterize these layers, we calculate the two-dimensional radial
distribution functions (2d rdf, $g_2(r)$) for the centers of mass
of the molecules within the layers.

All simulations are performed at $T=300$ K.

All simulations were performed using the LAMMPS simulation package
\cite{lammps}.

\section{Results and Discussion}

\subsection{Bulk PFB}

\begin{figure}

\includegraphics[width=8cm, height=6cm]{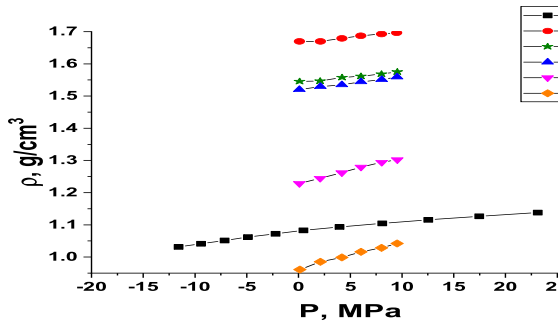}%

\caption{\label{bulk-compare}  A comparison of the equation of
state of bulk PFB obtained in this work (TW) with literature data.
JM - Ref. \cite{pfb-4}, HH - Ref. \cite{hh}, SCR - Ref.
\cite{shin}, BS - Ref. \cite{pvdf-ff-0}, MOD-exp - Ref.
\cite{pvdf-ff}}
\end{figure}

We start the discussion from the behavior of bulk PFB in a wide
range of pressures. Figure \ref{bulk-compare} shows a comparison
of our data with the experimental results and the results from
simulations of other authors. One can see that compared with the
experimental data our results underestimate the density of PFB at
given pressure. Moreover, only the model of Ref. \cite{pvdf-ff}
gives the equation of state in reasonable agreement with
experiment, all other computational models either underestimate,
or overestimate the density at fixed pressure. Although our
results are not in very good agreement with the experimental
results, we believe that our simulations give reasonable
qualitative results. Moreover, this model is computationally cheap
and allows rapid calculations which is important for calculations
of systems of longer fluorocarbons. Based on this we believe that
our model can serve as an opening gambit for simulation of
fluorocarbons of different chain lengths in bulk and confinement.

\begin{figure}

\includegraphics[width=8cm, height=6cm]{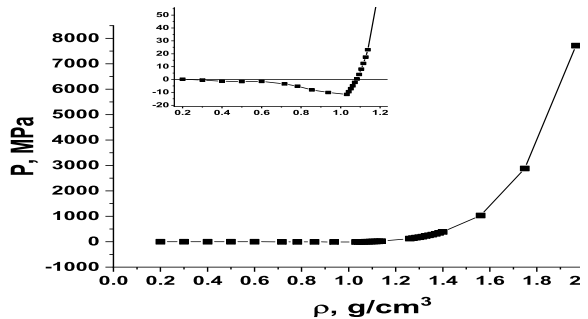}%

\caption{\label{bulk-eos} The equation of state of bulk PFB in a
wide range of pressures at $T=300$ K. The inset enlarges the part
of the plot at low densities.}
\end{figure}

Figure \ref{bulk-eos} shows the isotherm of PFB at $T=300$ K in a
wide range of densities. One can see that at the density below
$\rho=1.1$ $g/cm^3$ the pressure is negative which corresponds to
the appearance of gas - liquid transition in the molecular
simulation. The calculation of the boiling curve of PFB is beyond
the scope of the present paper. We just note that the density of
liquid at coexistence with gas is between the $\rho=1.03$ $g/cm^3$
and $\rho=1.14$ $g/cm^3$.

We discuss the structure of the liquid at two densities:
$\rho=1.4$ $g/cm^3$ and $\rho=1.97$ $g/cm^3$. The structure is
characterized by several partial RDFs: $C_1-C_1$, $C_1-C_4$,
$C_4-C_4$, which belong to the carbon chain of the molecules and
$F_3-F_3$, $F_3-H_3$, $H_3-H_3$ which show the distribution of
terminal fluorides and hydrogen atoms in the $CF_3$ and $CH_3$
groups.

\begin{figure}

\includegraphics[width=8cm, height=8cm]{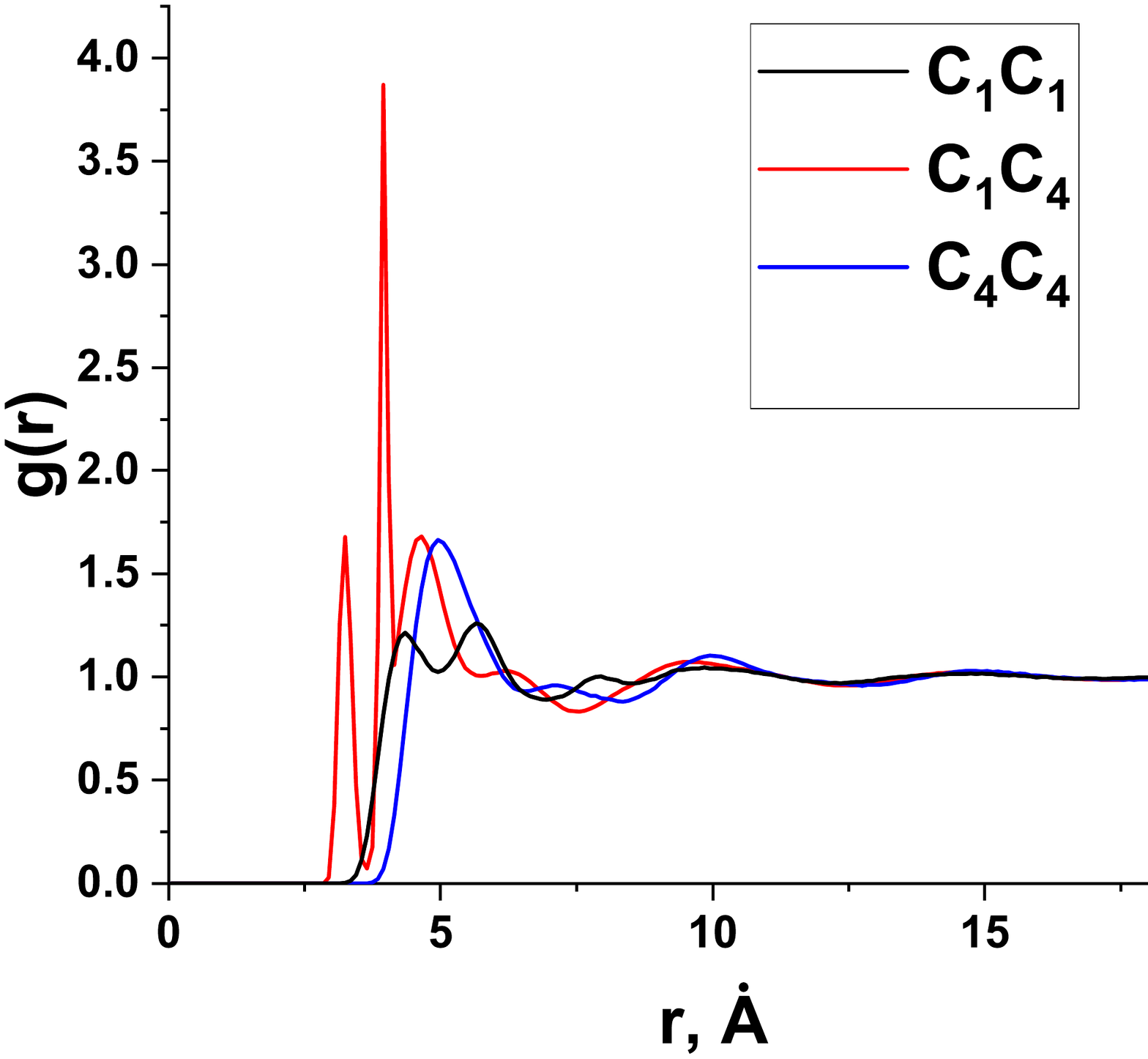}%

\includegraphics[width=8cm, height=8cm]{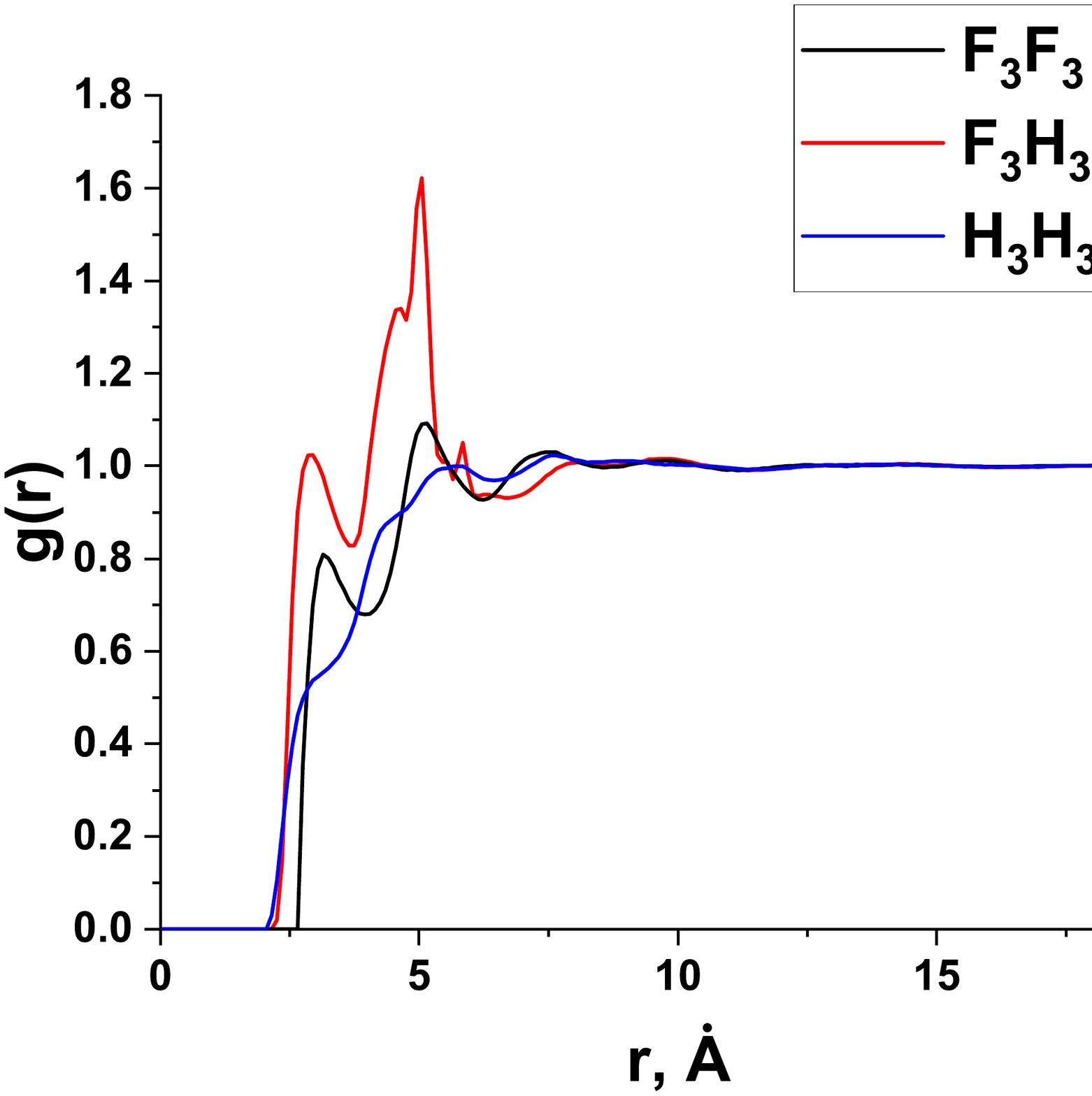}%

\caption{\label{rdf-56} The radial distribution functions of (a)
the carbon atoms and (b) the fluorines and hydrogens of PFB at
$\rho=1.4$ $g/cm^3$.}
\end{figure}

Figure \ref{rdf-56} shows the RDFs for density $\rho=1.4$
$g/cm^3$. One can see that the correlation between $C_1-C_1$ and
$C_4-C_4$ of different molecules is rather weak. However, the
correlation between $C_1$ and $C_4$ is much stronger, which means
that the molecules tend to orient by the $CH_3$ groups to the
$CF_3$ ones (Fig. \ref{rdf-56} (a)). The same conclusion can be
made from the correlations between the fluorine and hydrogen atoms
(Fig. \ref{rdf-56} (b)).

\begin{figure}

\includegraphics[width=8cm, height=8cm]{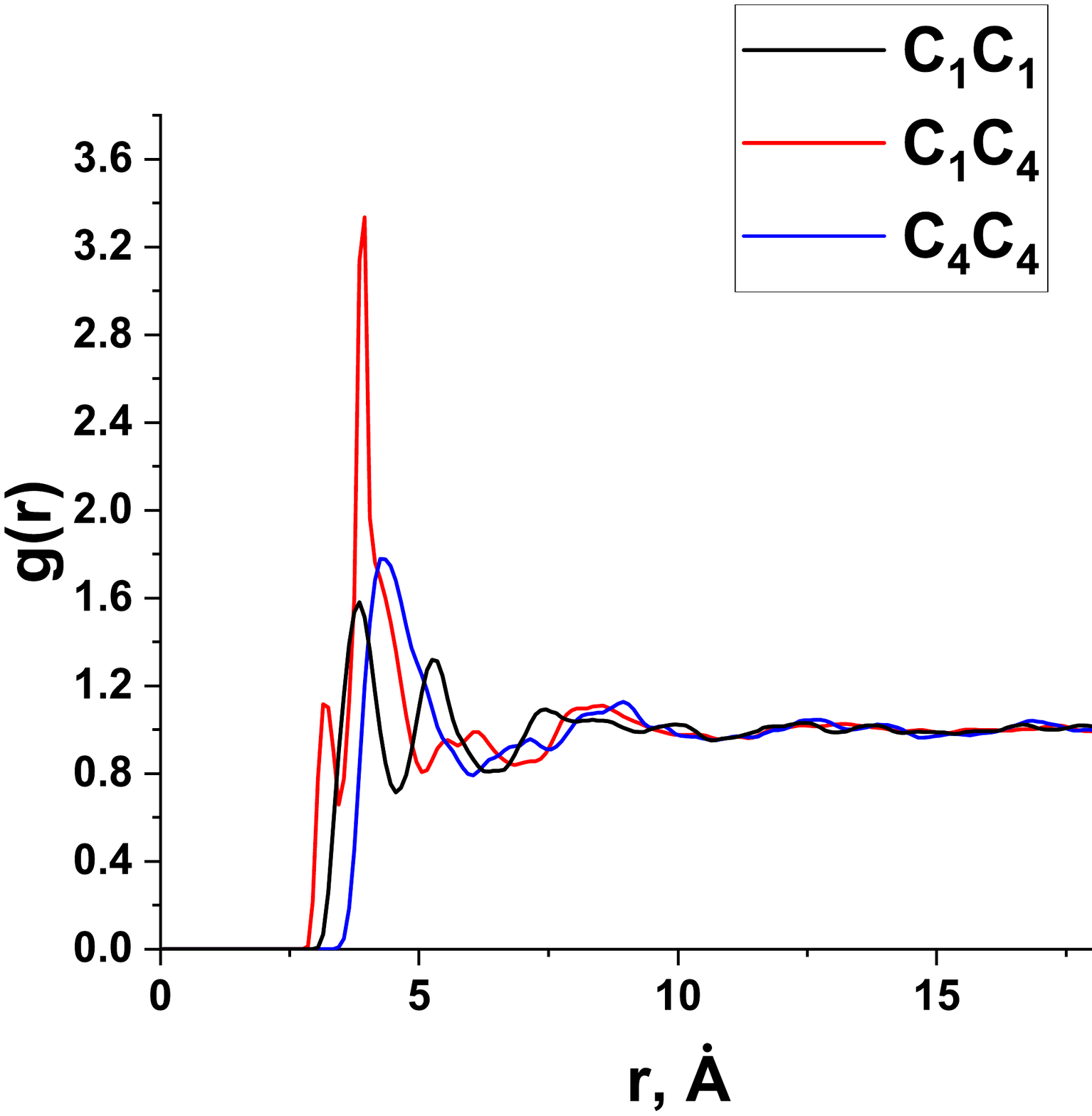}%

\includegraphics[width=8cm, height=8cm]{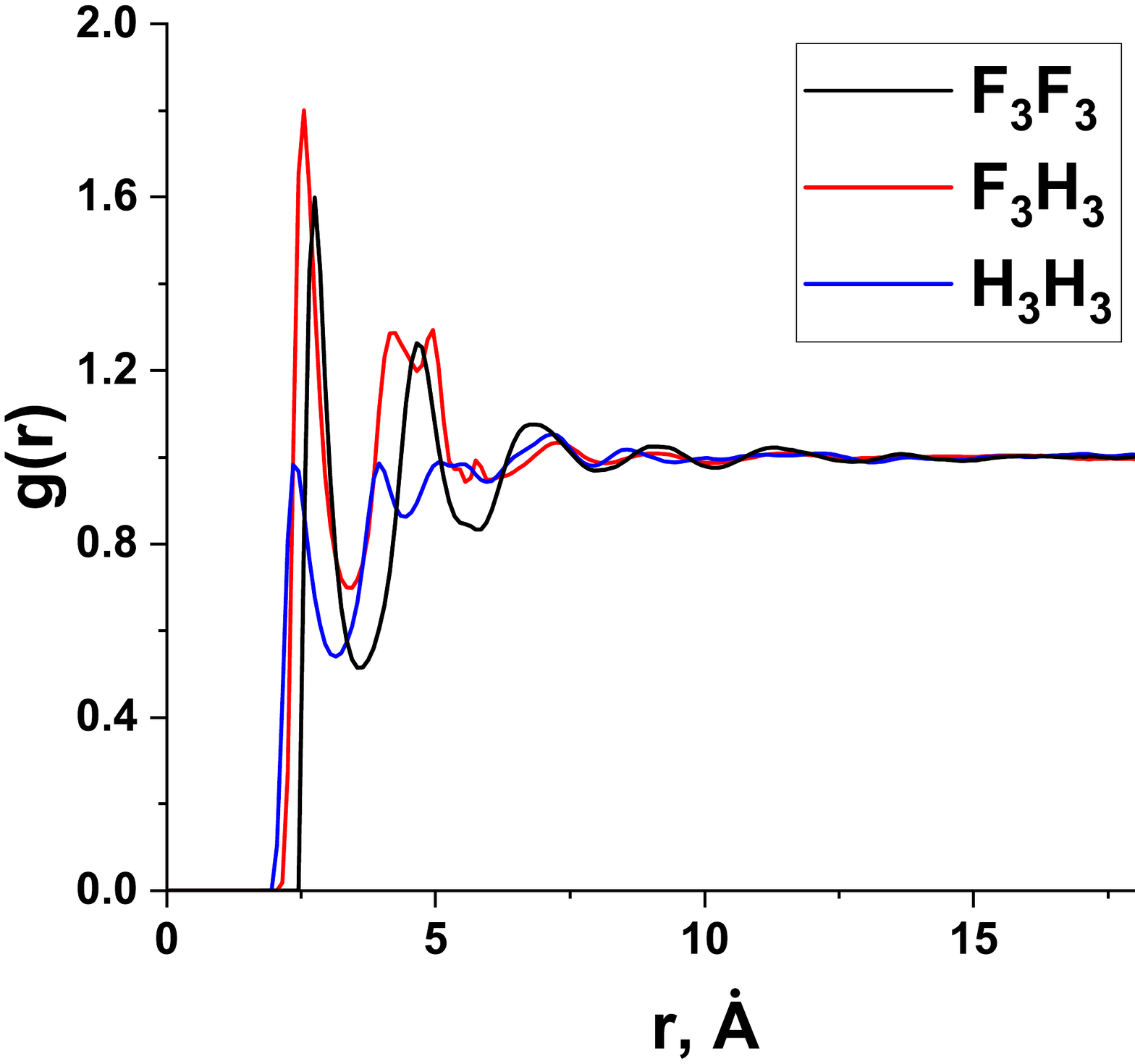}%

\caption{\label{rdf-50} The radial distribution functions of (a)
the carbon atoms and (b) the fluorines and hydrogens of PFB at
$\rho=1.97$ $g/cm^3$.}
\end{figure}

The density dependence on pressure starts to rapidly increase
after density $\rho=1.4$ $g/cm^3$. This might signal that the
liquid is close to the freezing line. However, even at density
$\rho=1.97$ $g/cm^3$ we do not observe any signatures of
crystallinity which can be seen from the partial RDFs given in
Fig. \ref{rdf-50} (a) and (b).

\subsection{Confined PFB}

\begin{figure}

\includegraphics[width=8cm, height=6cm]{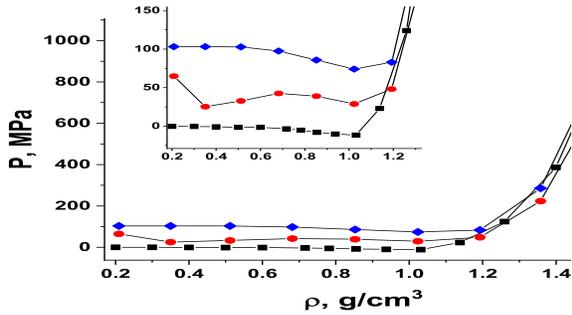}%

\caption{\label{eos-bulk-conf} A comparison of the equation of
state of bulk and confined PFB.}
\end{figure}

In this section we discuss the influence of confinement on the
thermodynamic and structural properties of PFB.  Figure
\ref{eos-bulk-conf} shows a comparison of the equation of state of
bulk and confined PFB. As usual, two components of the stress
tensor are considered in a slit pore:
$P_{||}=\frac{P_{xx}+P_{yy}}{2}$ which is parallel to the walls of
the pore and $P_{zz}$ which is perpendicular to the walls. One can
see that both $P_{||} $ and $P_{zz}$ demonstrate a van der Waals
loop which means the presence of gas-liquid transition in the
confined fluid. At the same time, this transition looks more
smeared in comparison with the bulk case. Apparently, the presence
of phase transition principally changes the structure of the
system, which we are going to study at the moment.

\begin{figure}

\includegraphics[width=6cm, height=5cm]{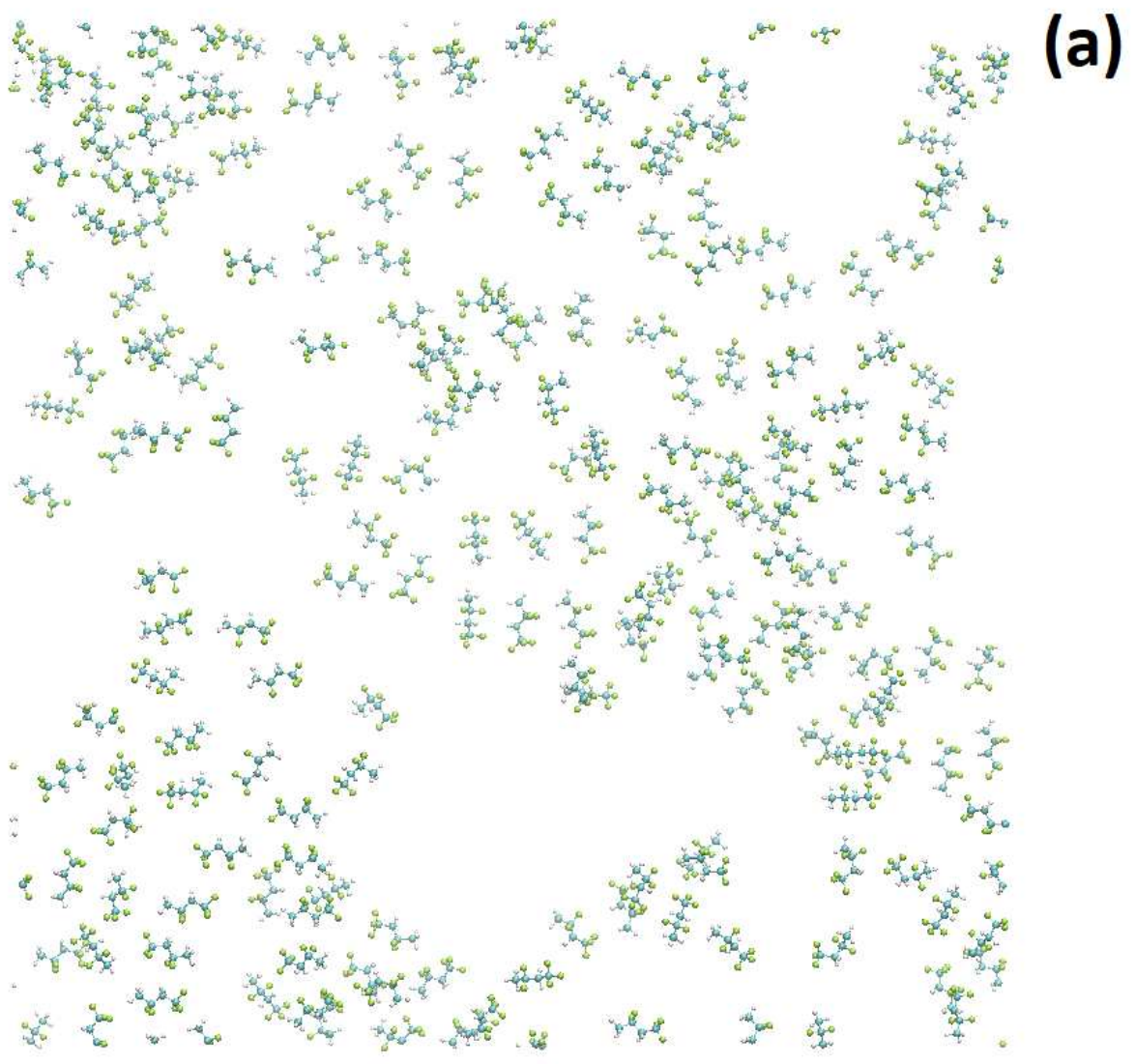}%

\includegraphics[width=6cm, height=4cm]{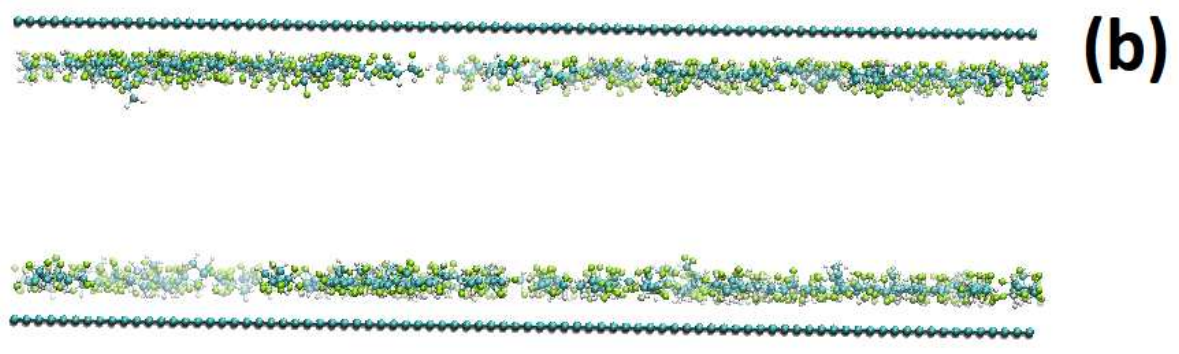}%

\includegraphics[width=8cm, height=6cm]{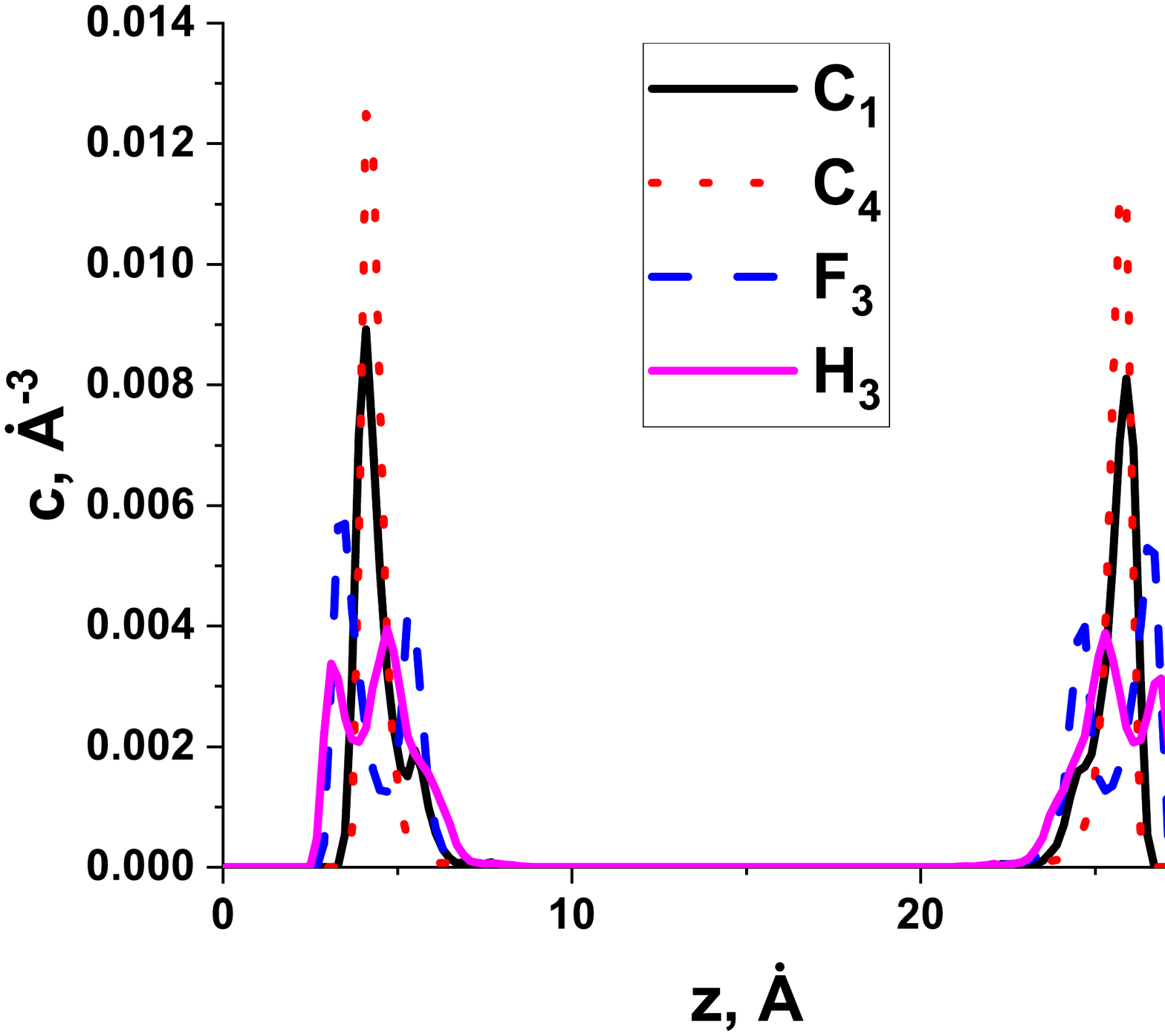}%

\includegraphics[width=8cm, height=6cm]{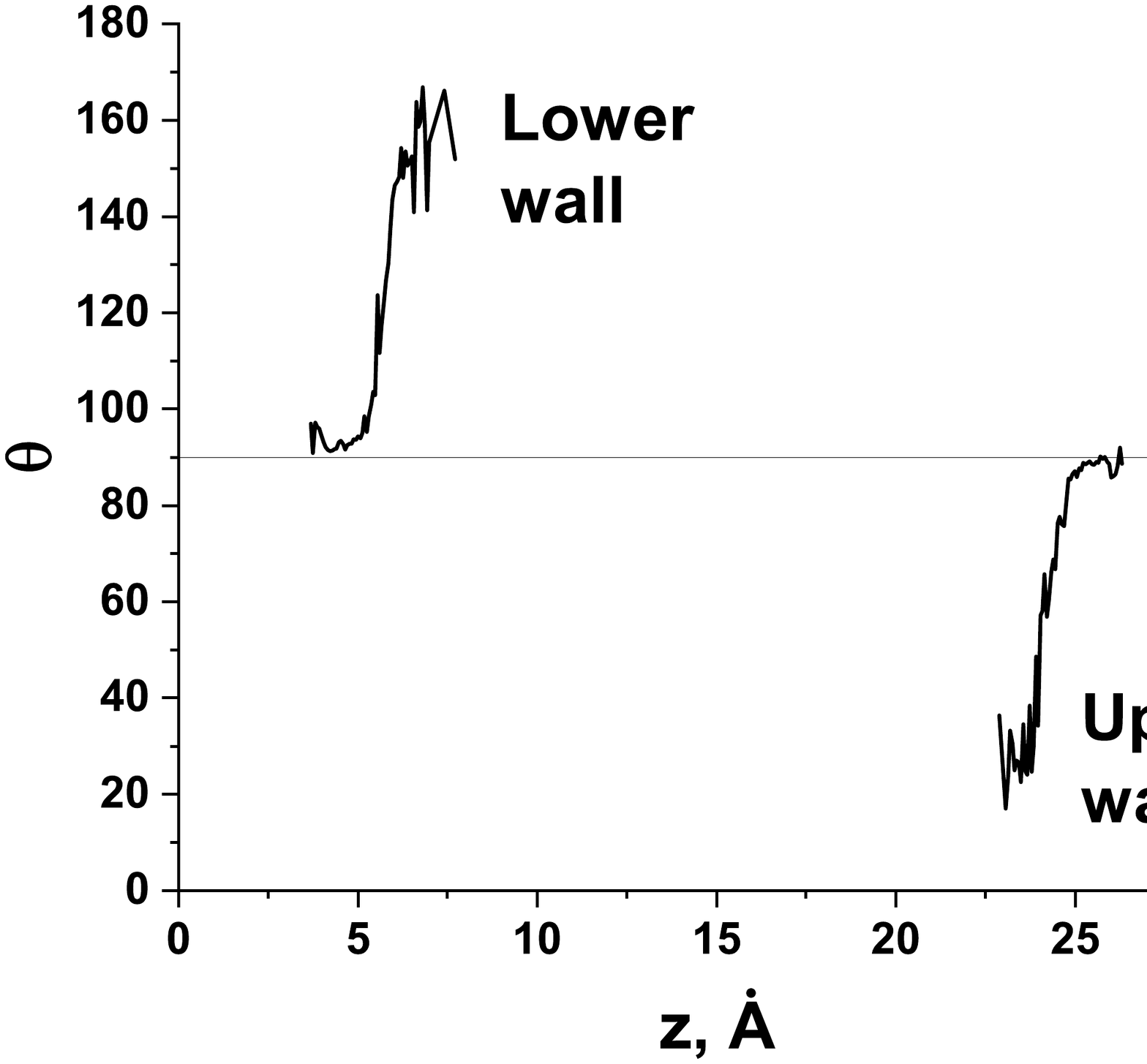}%

\caption{\label{200} (a) A top view of the system of 200 molecules
(density $\rho=0.21$ $g/cm^3$). The atoms of graphene are not
shown for clarity. (b) A side view of the same system. (c) The
distribution of number densities of several species in the same
system. For the $F_3$ and $H_3$ species the curves are divided by
3 in order to make them the same scale as $C_1$ and $C_4$. (d) The
distribution of angle $\theta$ along the z axis.}
\end{figure}

Figures \ref{200} (a) and (b) show a top and a side view of the
system of 200 PFB molecules in the graphene pore. From the top
view we see that the molecules are not condensed: we observe some
cavities in the system, which is characteristic of a gas-liquid
two phase region. The side view shows that the molecules strongly
stick to two walls. No exchange of molecules from the top wall to
the bottom one or vice versa is observed.

Figure \ref{200} (c) shows the number density (concentration)
distribution along the z-axis for several types of atoms: $C_1$
(the carbon atom in the $CH_3$ group), $C_4$ (the carbon atom in
the $CF_3$ group), $F_3$ (the fluorine atom in the $CF_3$ group)
and $H_3$ (the hydrogen atom in the $CH_3$ group). The $C_4$ atoms
approach a bit closer to the walls than the $C_1$ ones.

In order to evaluate the orientation of the molecules next to the
graphene walls in more detail we calculate the distribution of
angle $\theta$ introduced in the "System and Methods" section. The
results are shown in Fig. \ref{200} (d). One can see that next to
the walls $\theta$ is close to 90 degrees, which means that the
molecules lie almost parallel to the walls. This result is
consistent with the fact that the location of the main peak of
$C_1$ and $C_4$ coincide. At the same time, as the molecules
deviate from the wall the angle deviates from a right angle in
such a way that the $CF_3$ group appears to be further from the
wall than the $CH_3$ one. This corresponds to an obtuse angle for
the upper wall and to an acute angle for the lower one. For the
reasons of symmetry, the curves for the upper and lower walls are
symmetric with respect to the line of 90 degrees.



\begin{figure}

\includegraphics[width=6cm, height=5cm]{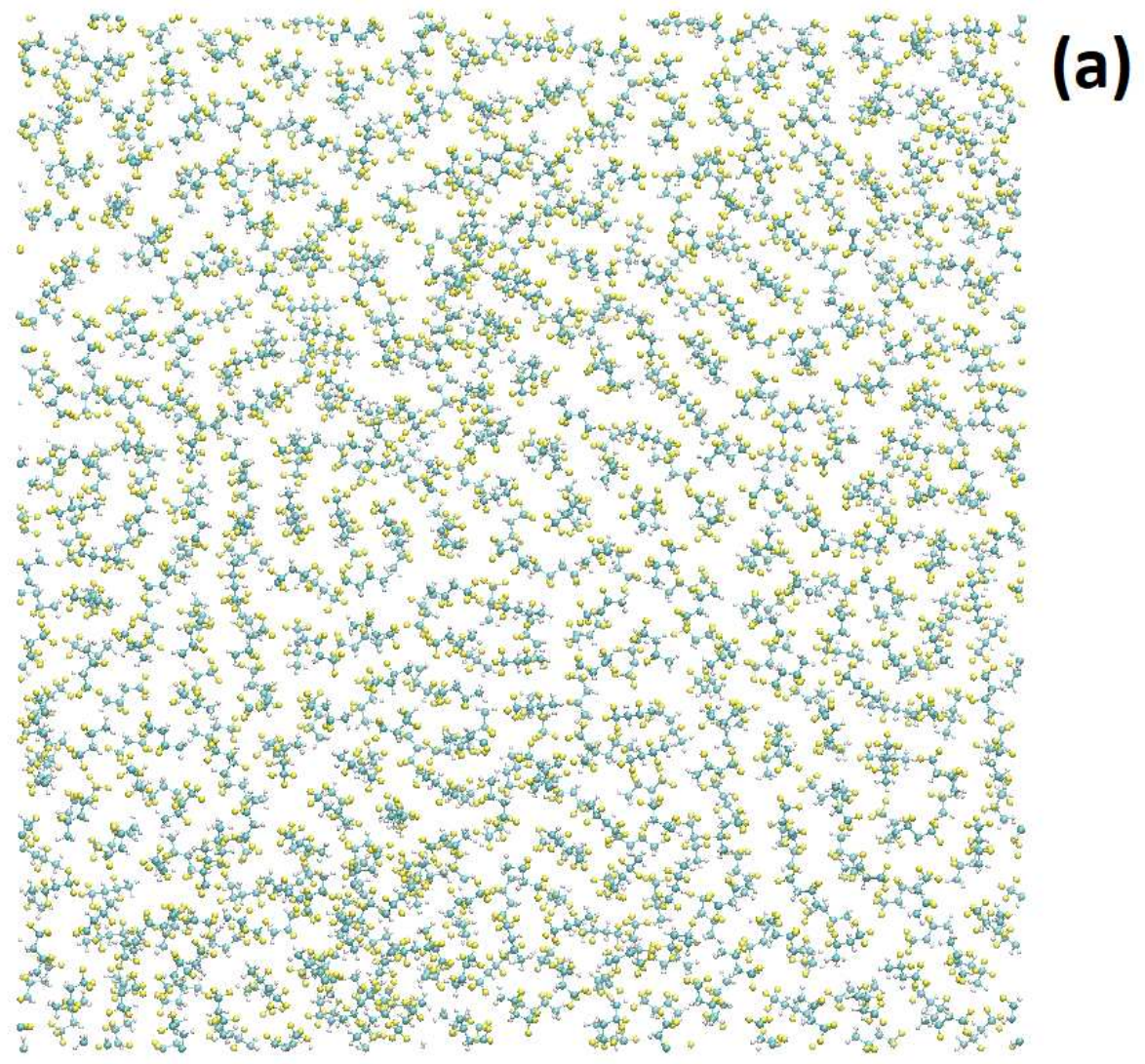}%

\includegraphics[width=6cm, height=4cm]{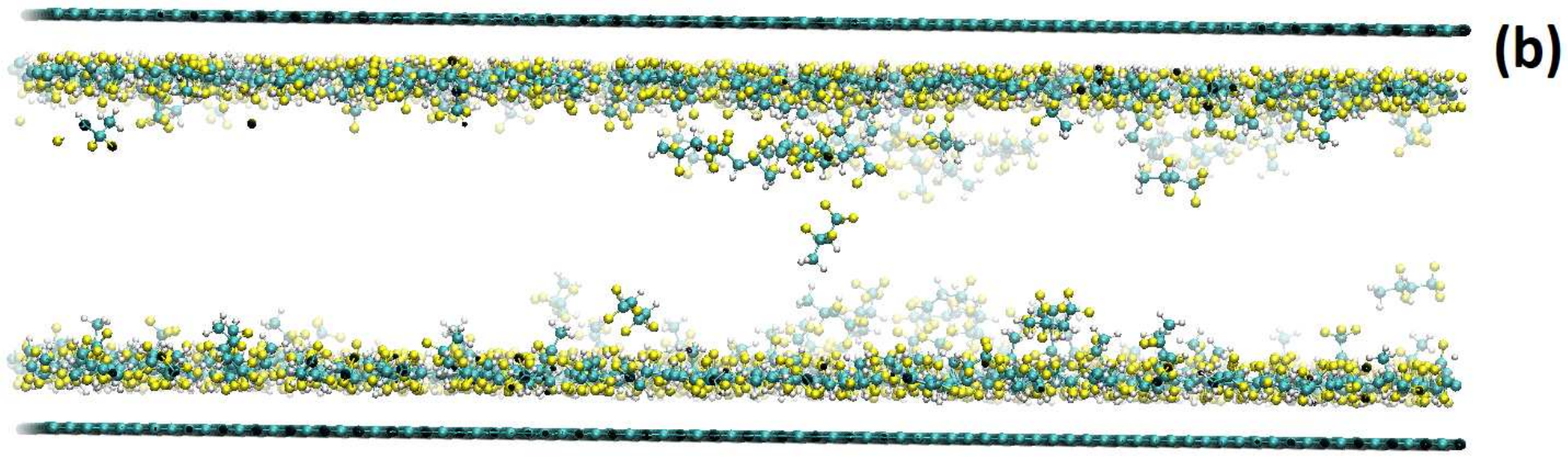}%

\includegraphics[width=8cm, height=6cm]{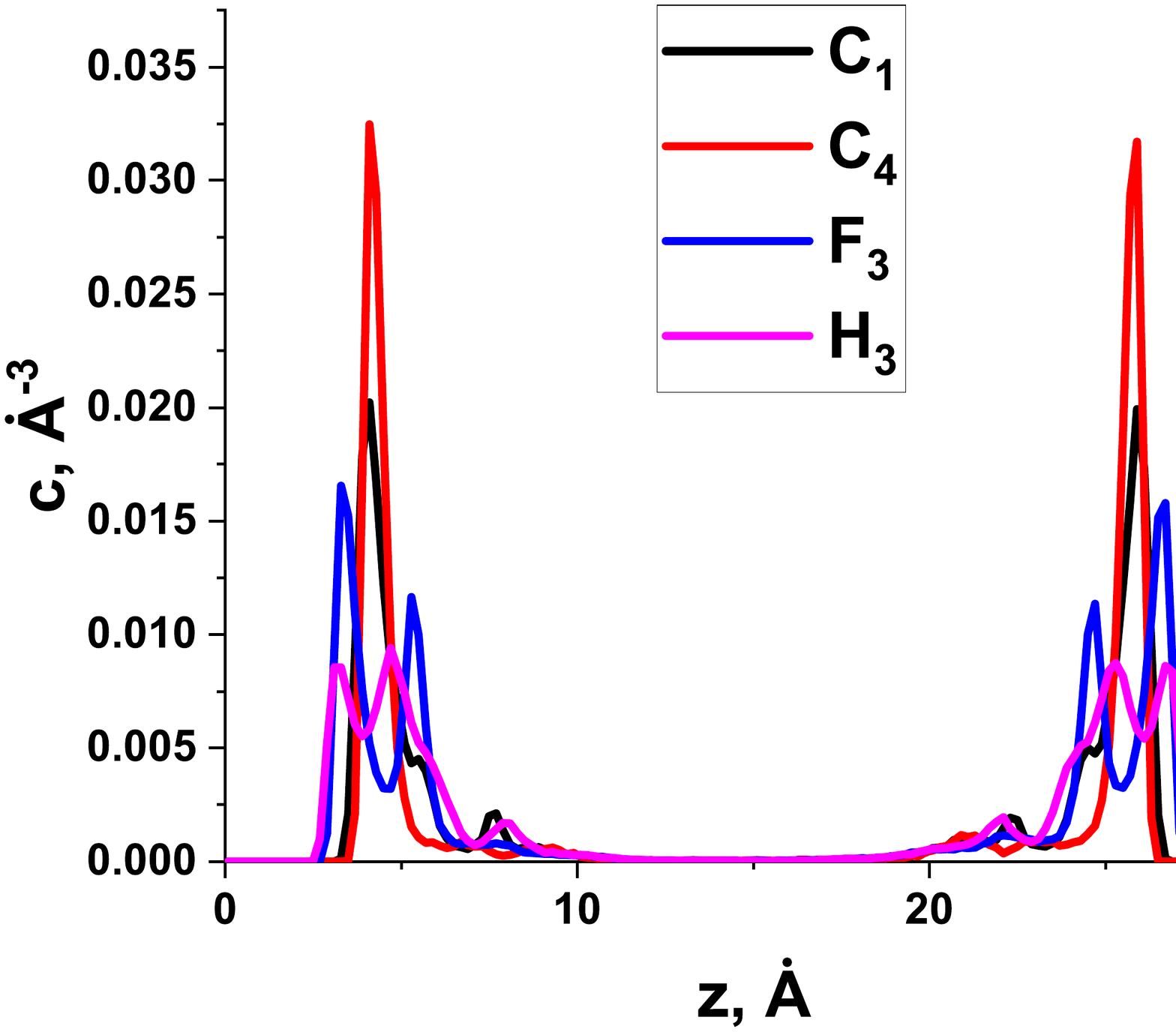}%

\includegraphics[width=8cm, height=6cm]{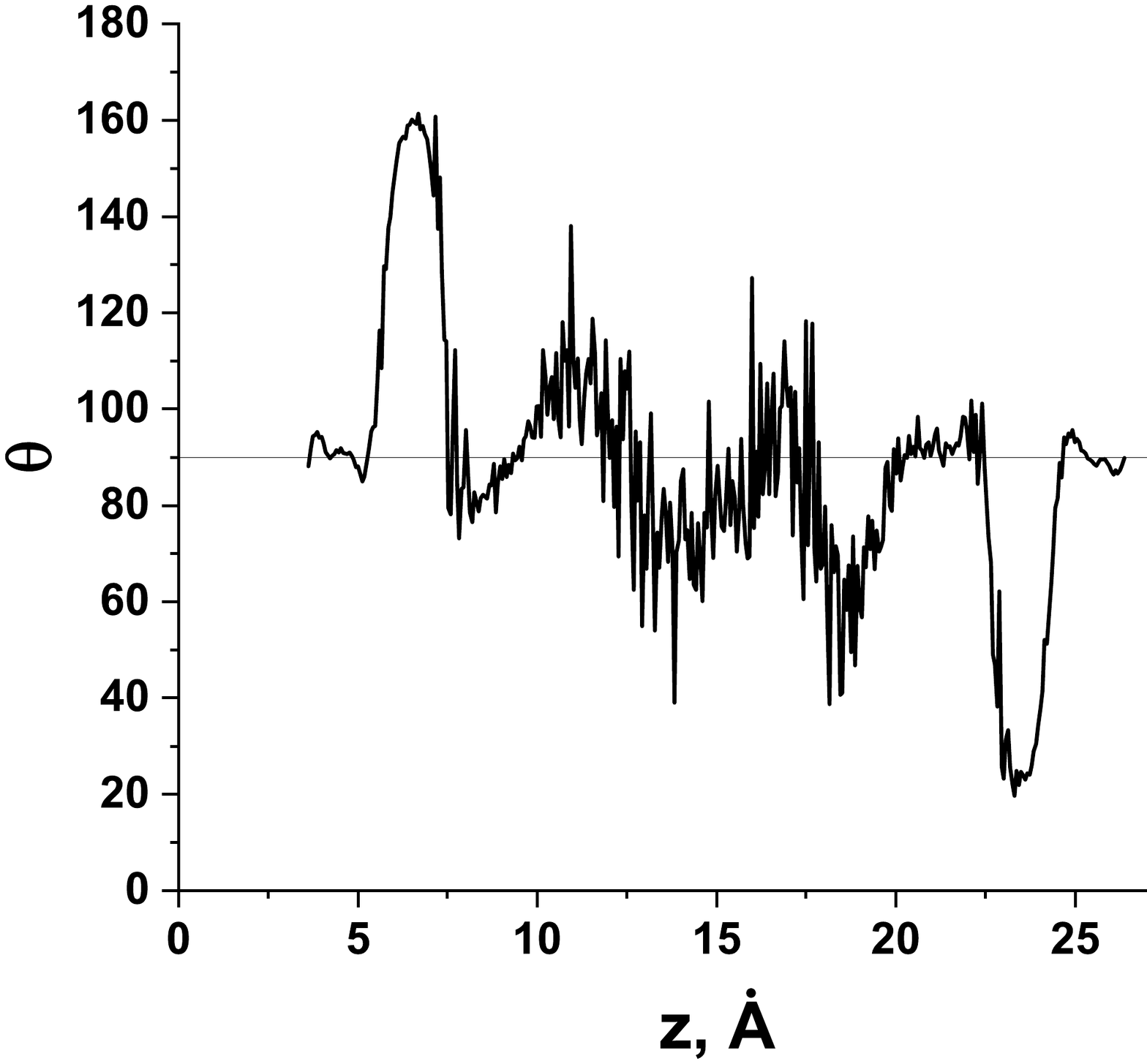}%

\caption{\label{600} (a) A top view of the system of 600 molecules
(density $\rho=0.51$ $g/cm^3$). The atoms of graphene are not
shown for clarity. (b) A side view of the same system. (c) The
distribution of number densities of several species in the same
system. For the $F_3$ and $H_3$ species the curves are divided by
3 in order to make them the same scale as $C_1$ and $C_4$. (d) The
distribution of angle $\theta$ along the z axis.}
\end{figure}

When the number of molecules is increased, they adsorb on the
walls up to the moment when all available space on the walls is
occupied (see Fig. \ref{pvdf-mol} in the Supplementary materials
for the system of 400 PFB molecules in the pore). After that the
molecules go into the free space between the walls. Figure
\ref{600} (a) shows a top view of the system of 600 molecules of
PFB in the pore. One can see that no free space on the walls is
available to be occupied by the molecules. From Fig. \ref{600} (b)
one can see that only a few molecules are pushed into the middle
of the pore. This is confirmed by the local density distribution
of the species shown in Fig. \ref{600} (c).

Figure \ref{600} (d) shows the angle distribution of the system of
600 molecules of PFB in the pore. One can see that next to the
wall the distribution behaves as in the case of 200 molecules:
they lie flat on the walls. As the molecules go further from the
walls they pass a maximum (minimum) at 160 (20) degrees and
starting from the distance of 10 $\AA$ from the walls the
orientation fluctuates around a right angle.

\begin{figure}

\includegraphics[width=6cm, height=5cm]{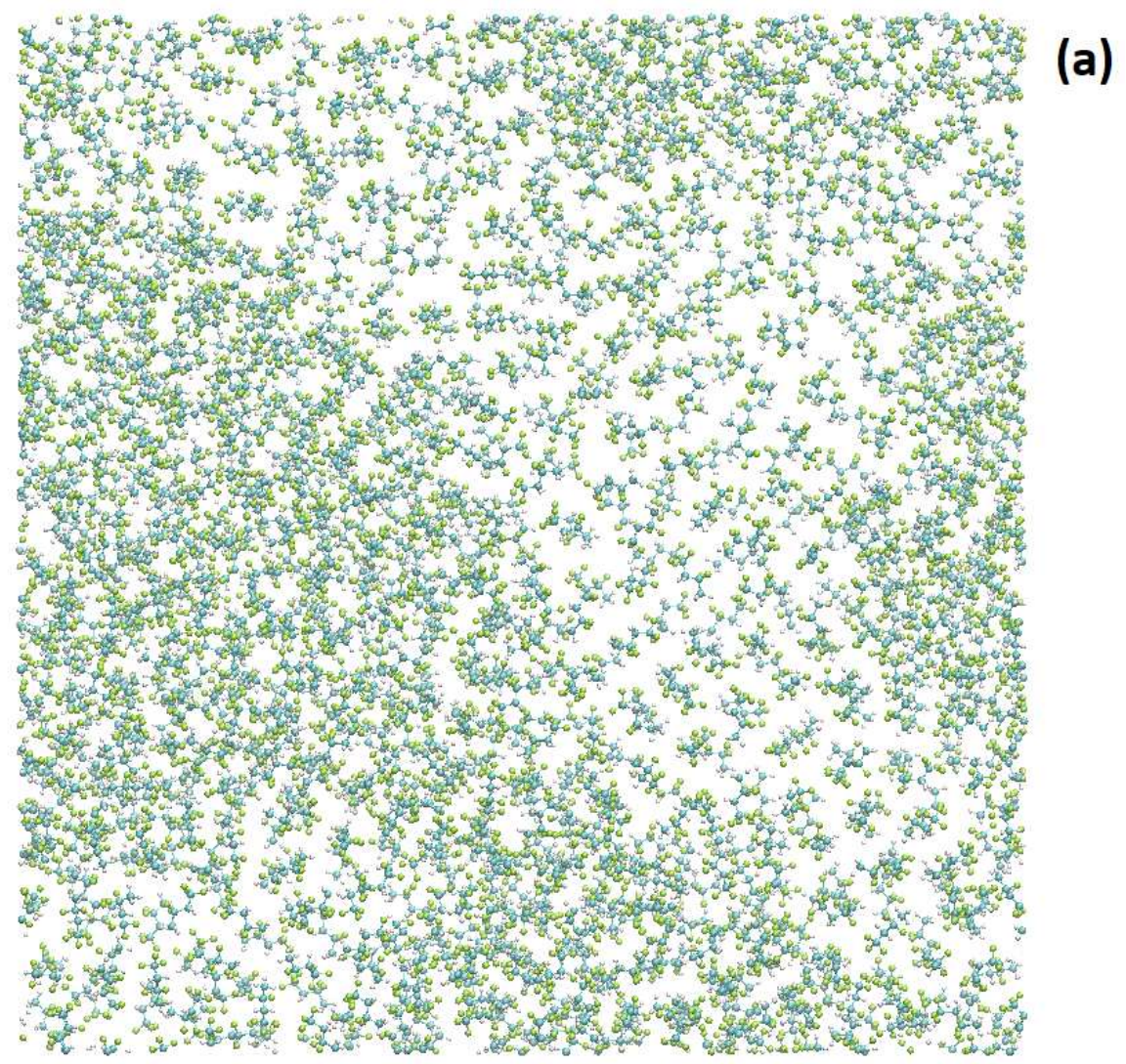}%

\includegraphics[width=6cm, height=4cm]{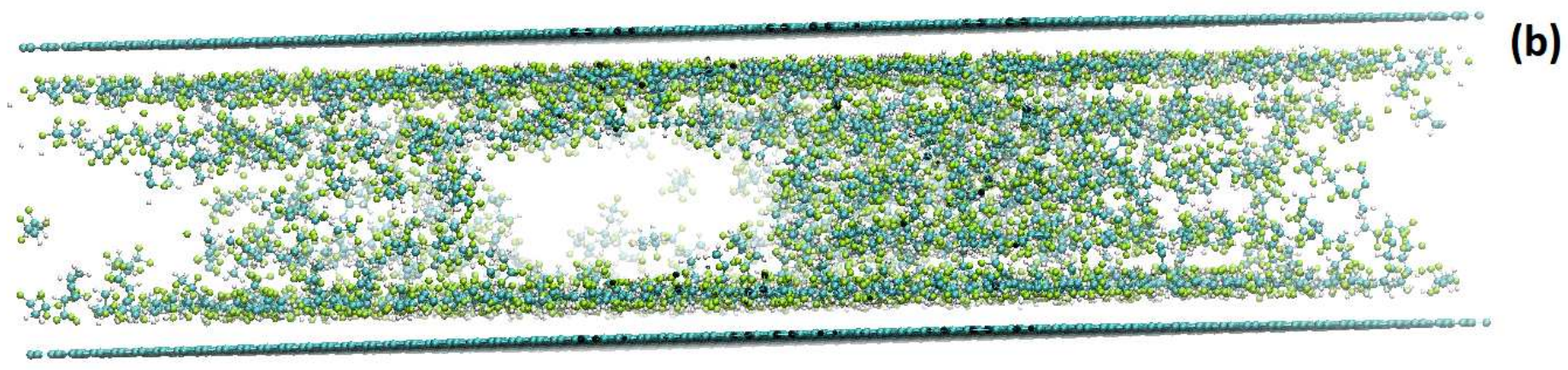}%

\includegraphics[width=8cm, height=6cm]{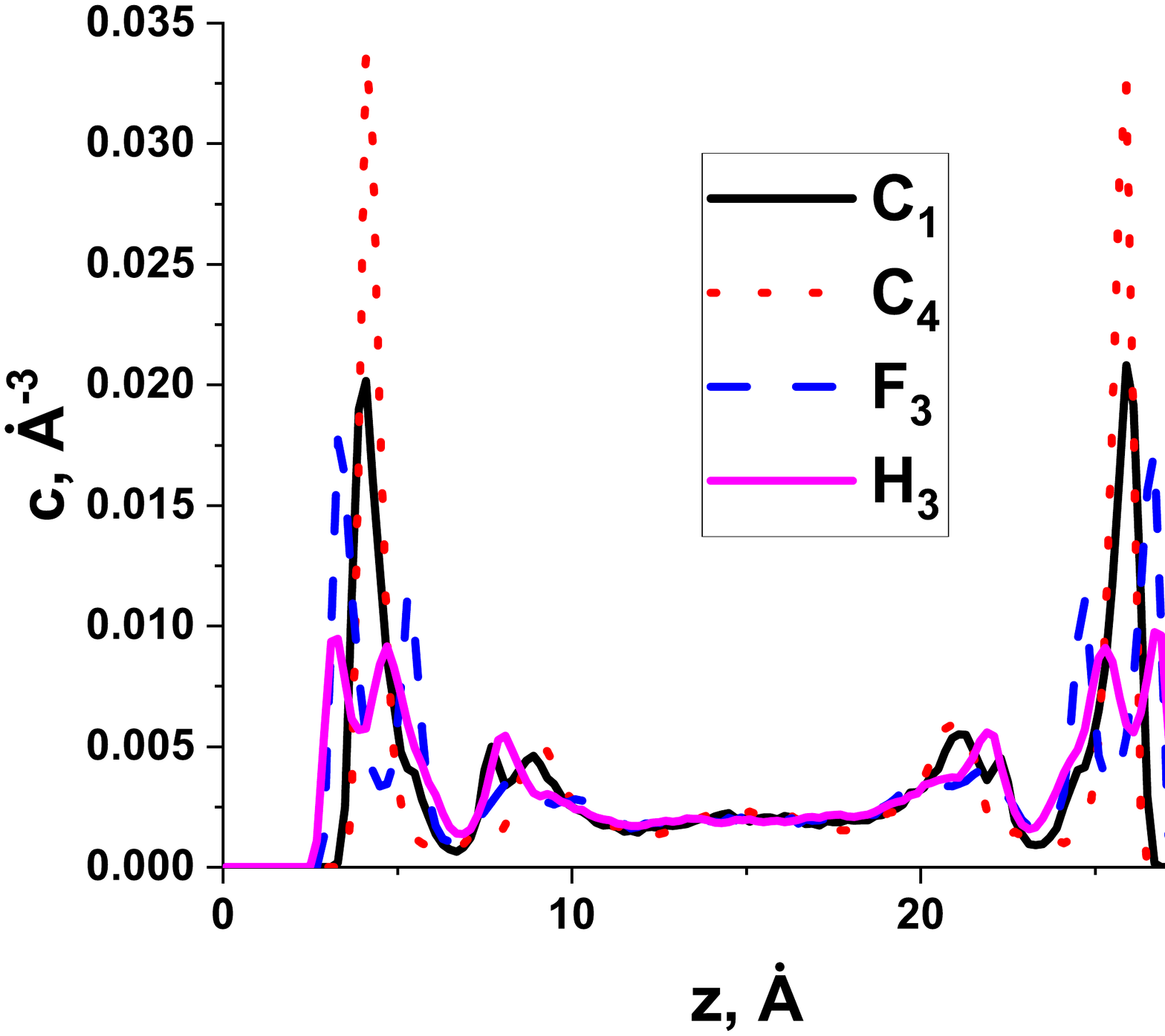}%

\includegraphics[width=8cm, height=6cm]{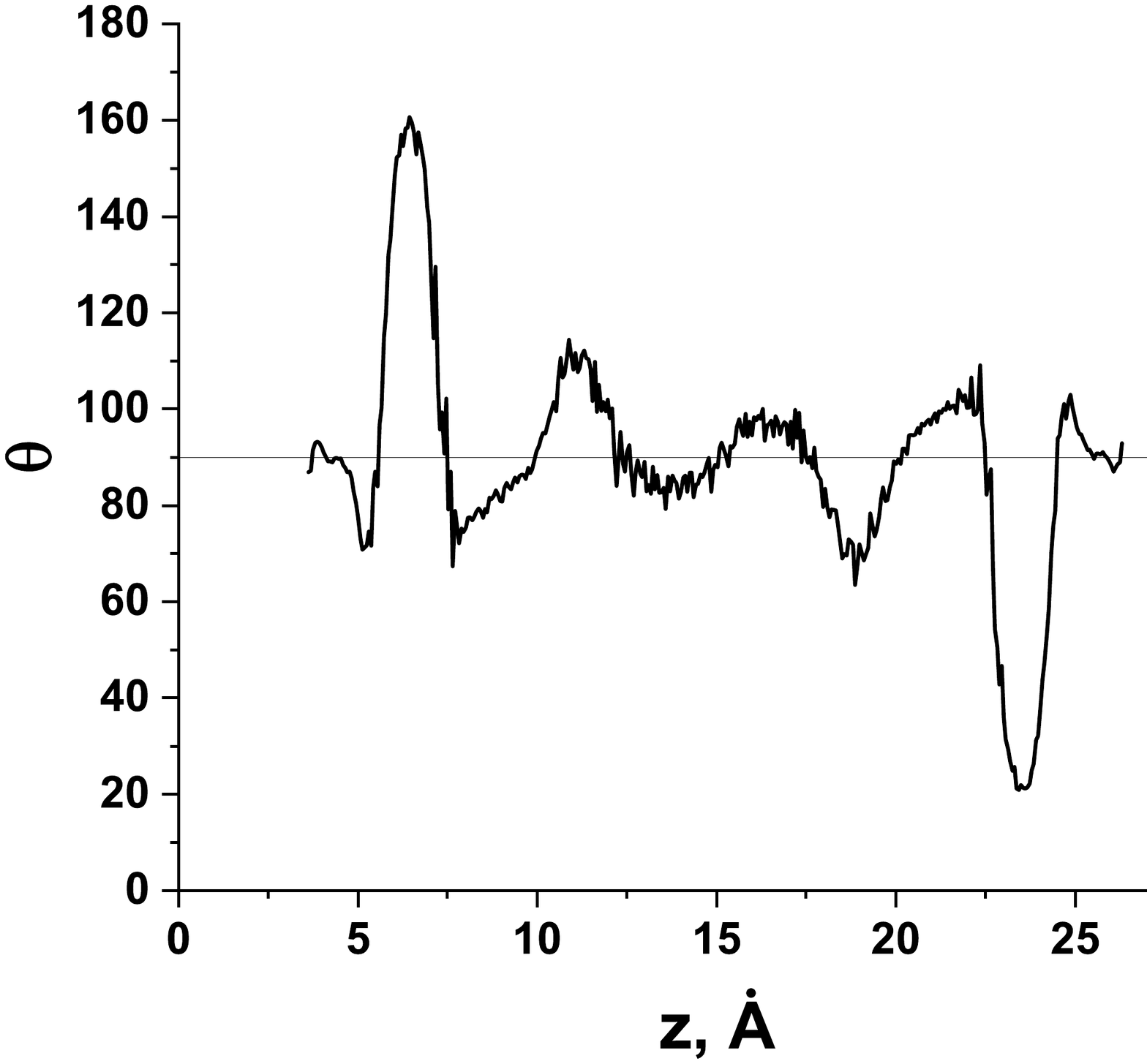}%

\caption{\label{1000} (a) A top view of the system of 1000
molecules (density $\rho=0.85$ $g/cm^3$). The atoms of graphene
are not shown for clarity. (b) A side view of the same system. (c)
The distribution of number densities of several species in the
same system. For the $F_3$ and $H_3$ species the curves are
divided by 3 in order to make them the same scale as $C_1$ and
$C_4$. (d) The distribution of angle $\theta$ along the z axis.}
\end{figure}

We proceed with investigation of the system with density
$\rho=0.85$ $g/cm^3$ ($N=1000$). Figures \ref{1000} (a) and (b)
show the top and side views of this system, respectively. From
these pictures one can see that some parts of the system are
condensed, while some others are dilute. This corresponds to the
gas - liquid two-phase region, which has been previously shown on
the EoS of the system (see Fig. \ref{eos-bulk-conf}). Note that
the bulk system is also in the two-phase region at this density.
Figure \ref{1000} (c) demonstrates the density distribution along
the z axis. Comparing with the previous densities we observe that
the density distribution of $C_4$ and $C_1$ demonstrates two peaks
(at 4.1 and 9.1 $\AA$) from each wall and three peaks (at 4.1, 7.7
and 9.1 $\AA$) from each wall, respectively, which means that the
system becomes more ordered.

The distribution of angle $\theta$ demonstrates a modulated
structure in the whole volume of the pore. The orientation of the
molecules within the layers next to the walls does not change
compared with the cases of smaller densities. The molecules of the
second layer demonstrate an orientation of 110 (the upper wall) or
70 (the lower wall) degrees to the walls.

\begin{figure}

\includegraphics[width=6cm, height=5cm]{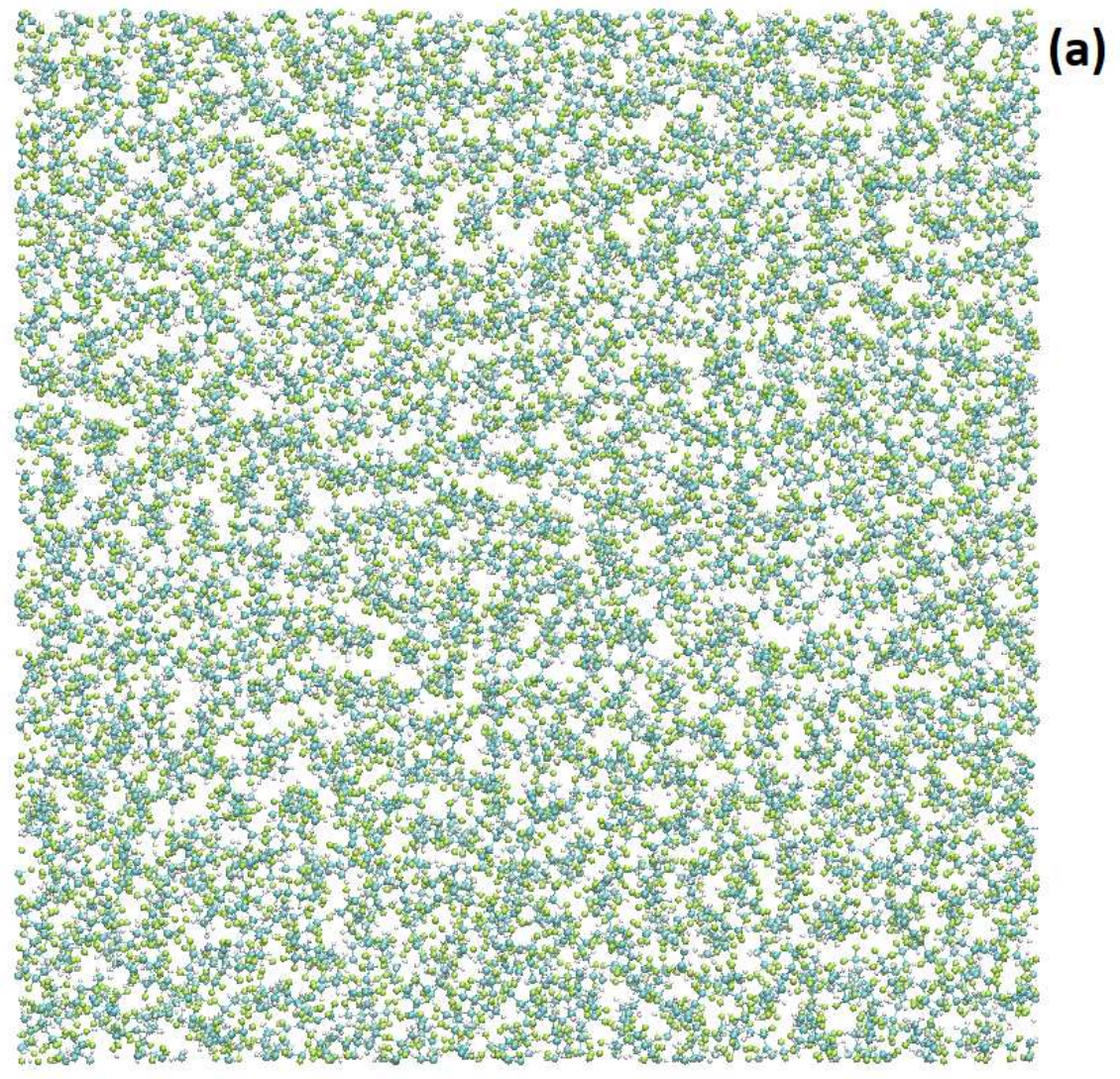}%

\includegraphics[width=6cm, height=4cm]{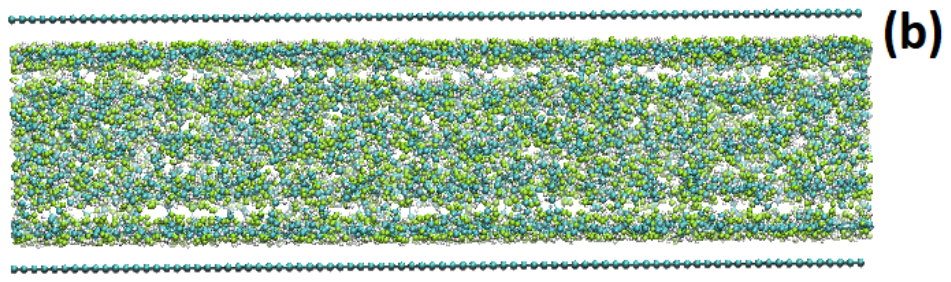}%

\includegraphics[width=8cm, height=6cm]{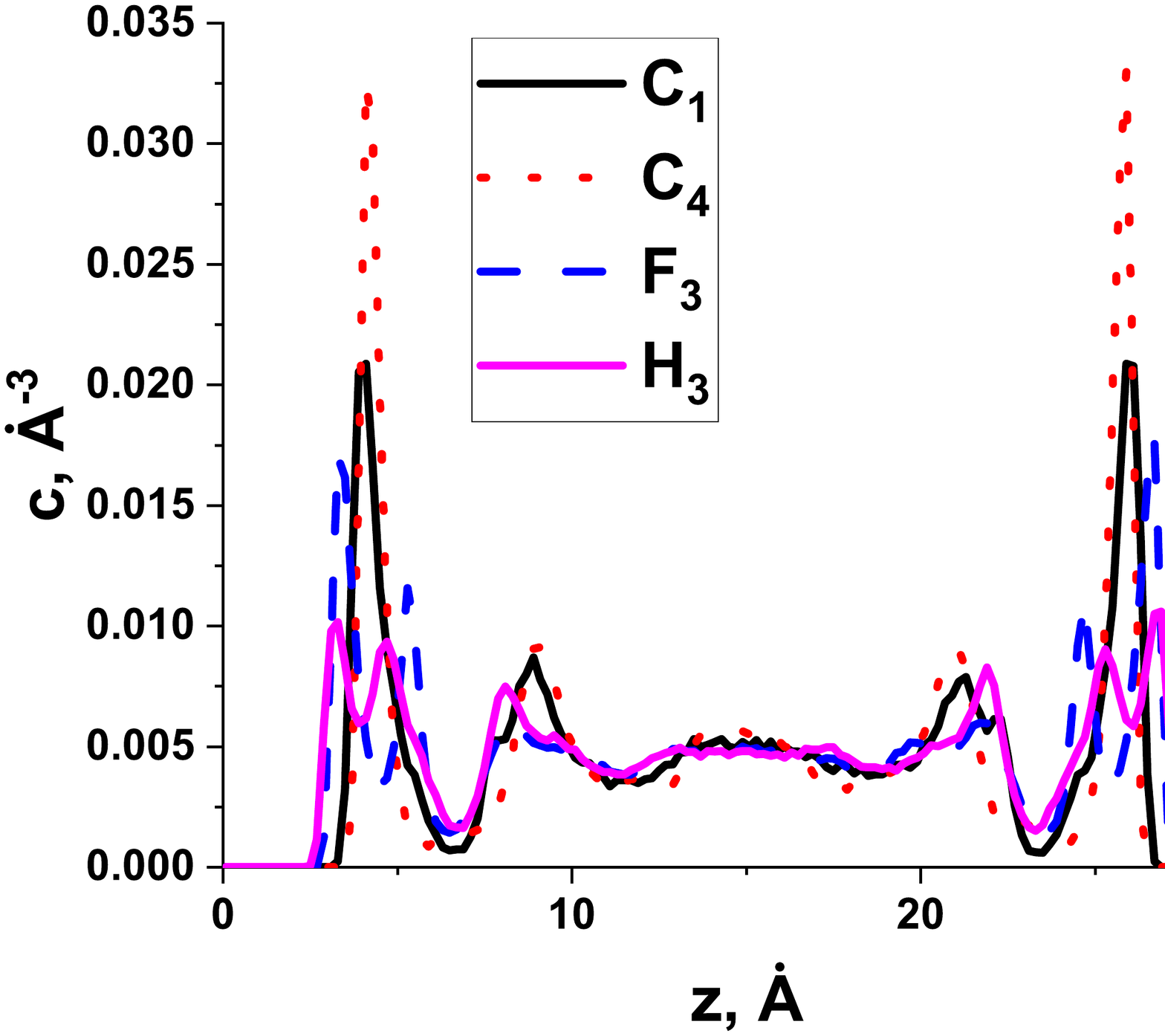}%

\includegraphics[width=8cm, height=6cm]{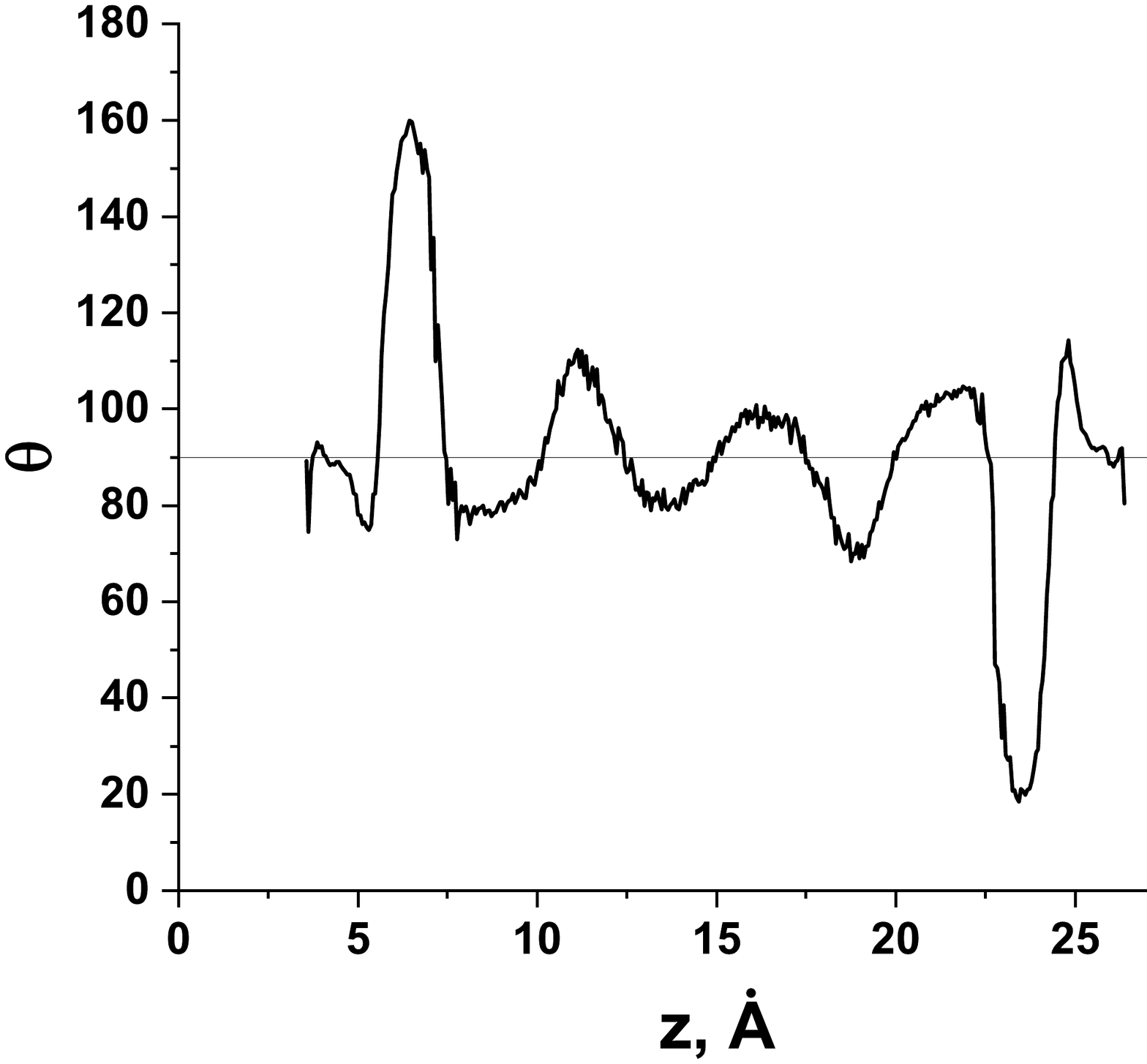}%

\caption{\label{1400} (a) A top view of the system of 1400
molecules (density $\rho=1.19$ $g/cm^3$). The atoms of graphene
are not shown for clarity. (b) A side view of the same system. (c)
The distribution of number densities of several species in the
same system. For the $F_3$ and $H_3$ species the curves are
divided by 3 in order to make them the same scale as $C_1$ and
$C_4$. (d) The distribution of angle $\theta$ along the z axis.}
\end{figure}

Figure \ref{1400} shows the same data for the system with density
$\rho=1.19$ $g/cm^3$ ($N=1400$). As compared with the previous
case the first minimum of the density of the $C_1$ and $C_4$
species falls almost to a zero value (see Fig. \ref{1400} (c)),
i.e. the first layer appears to be almost isolated from the other
ones, which can be seen in the snapshots of the system (Fig.
\ref{1400} (a) and (b)). The distribution of angle $\theta$
becomes even more modulated (Fig. \ref{1400} (d)).

\begin{figure}

\includegraphics[width=6cm, height=5cm]{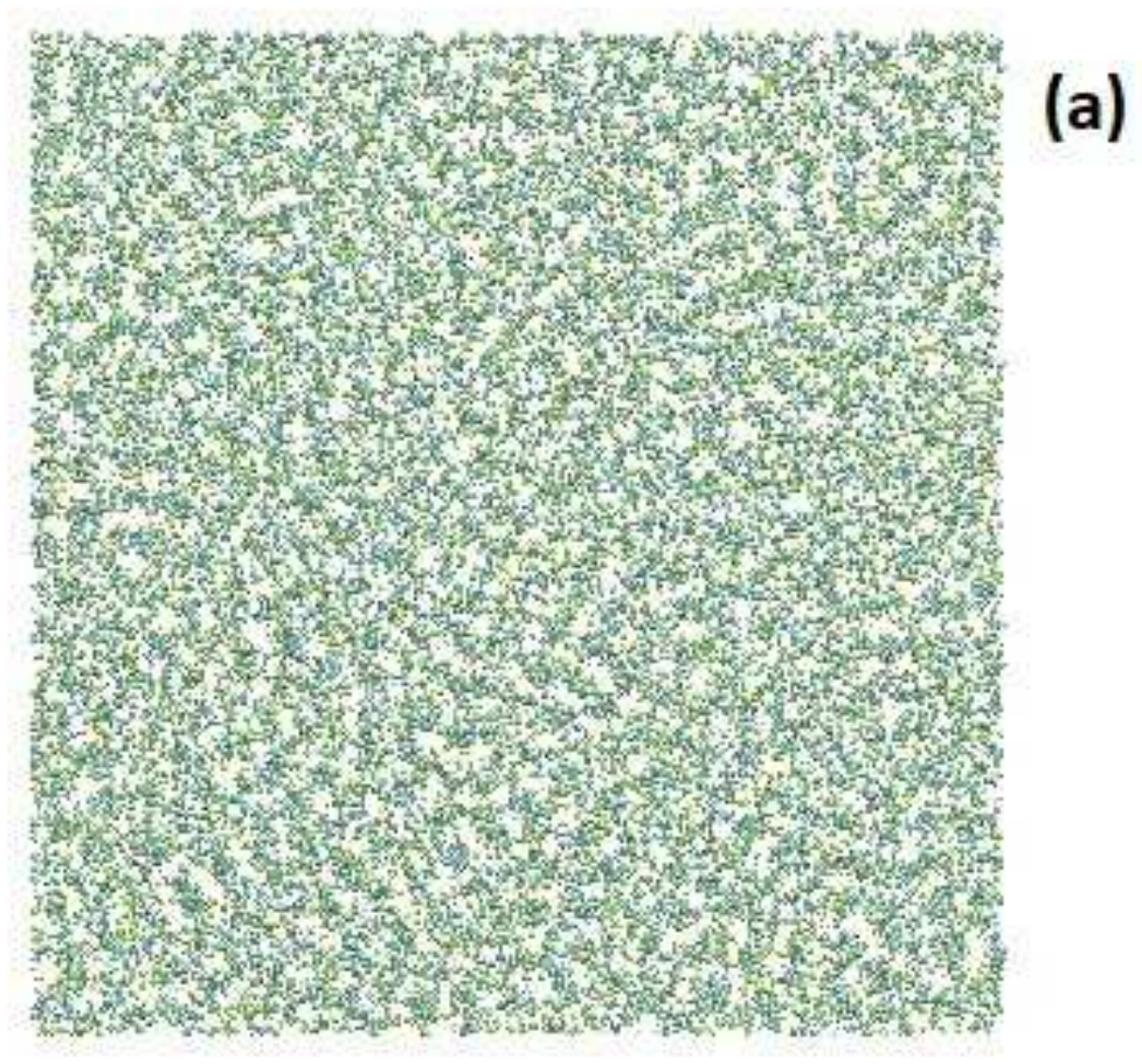}%

\includegraphics[width=6cm, height=4cm]{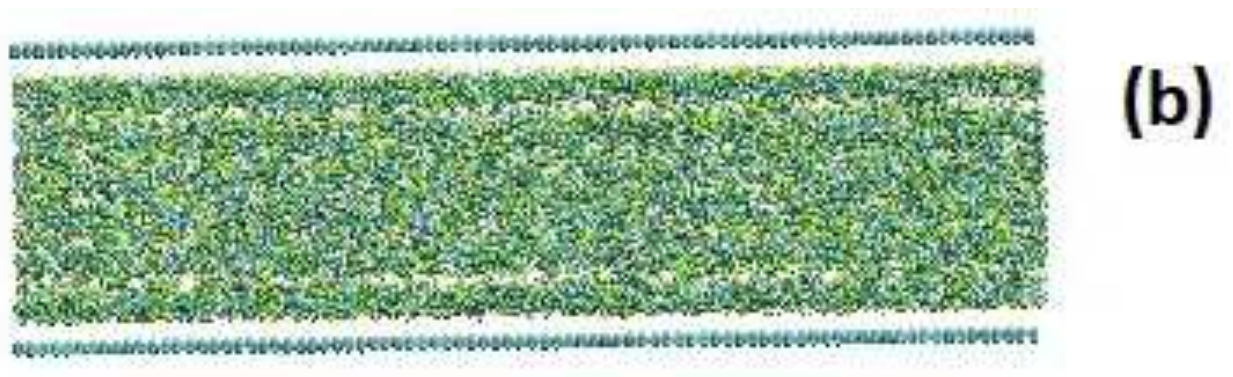}%

\includegraphics[width=8cm, height=6cm]{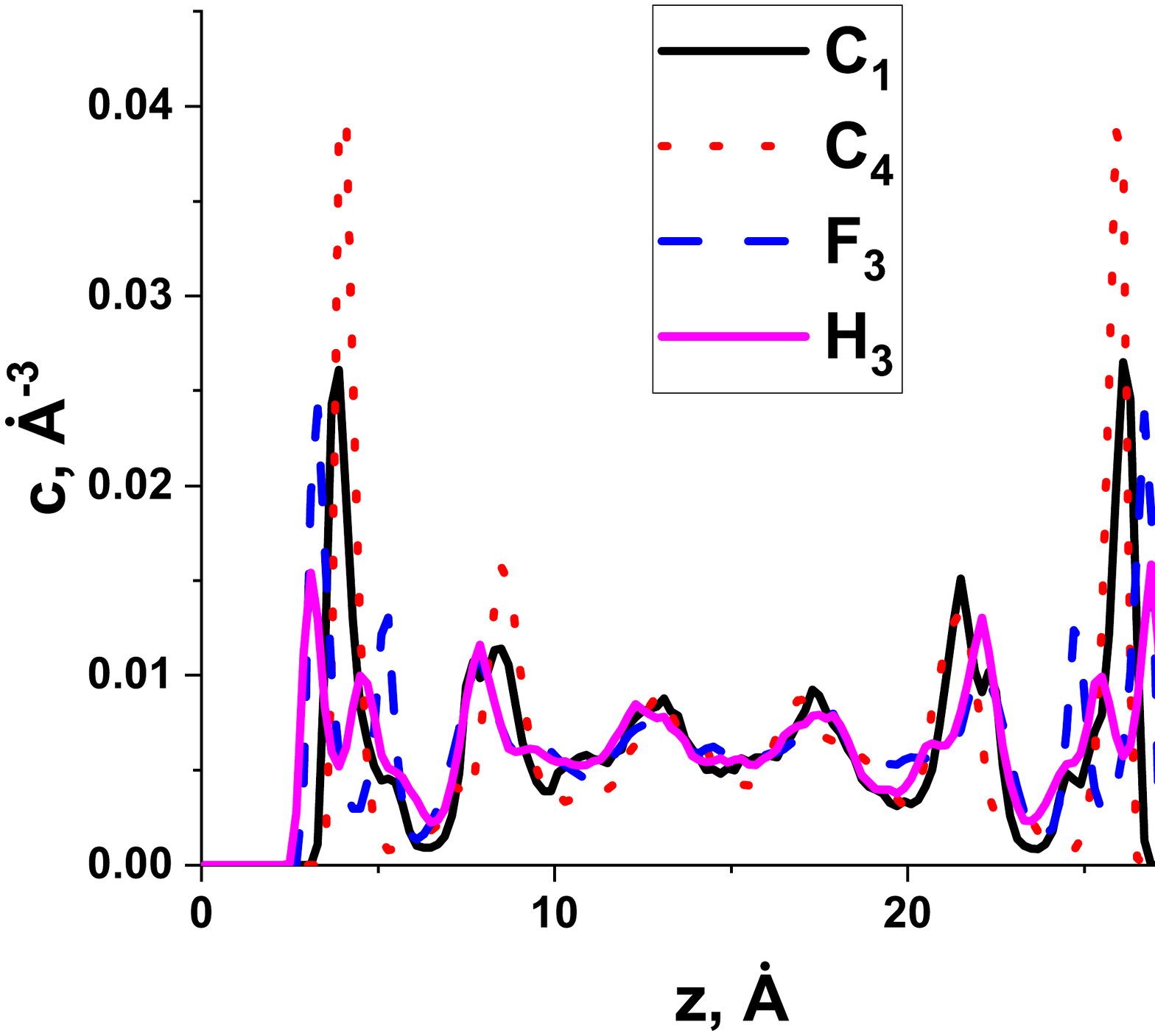}%

\includegraphics[width=8cm, height=6cm]{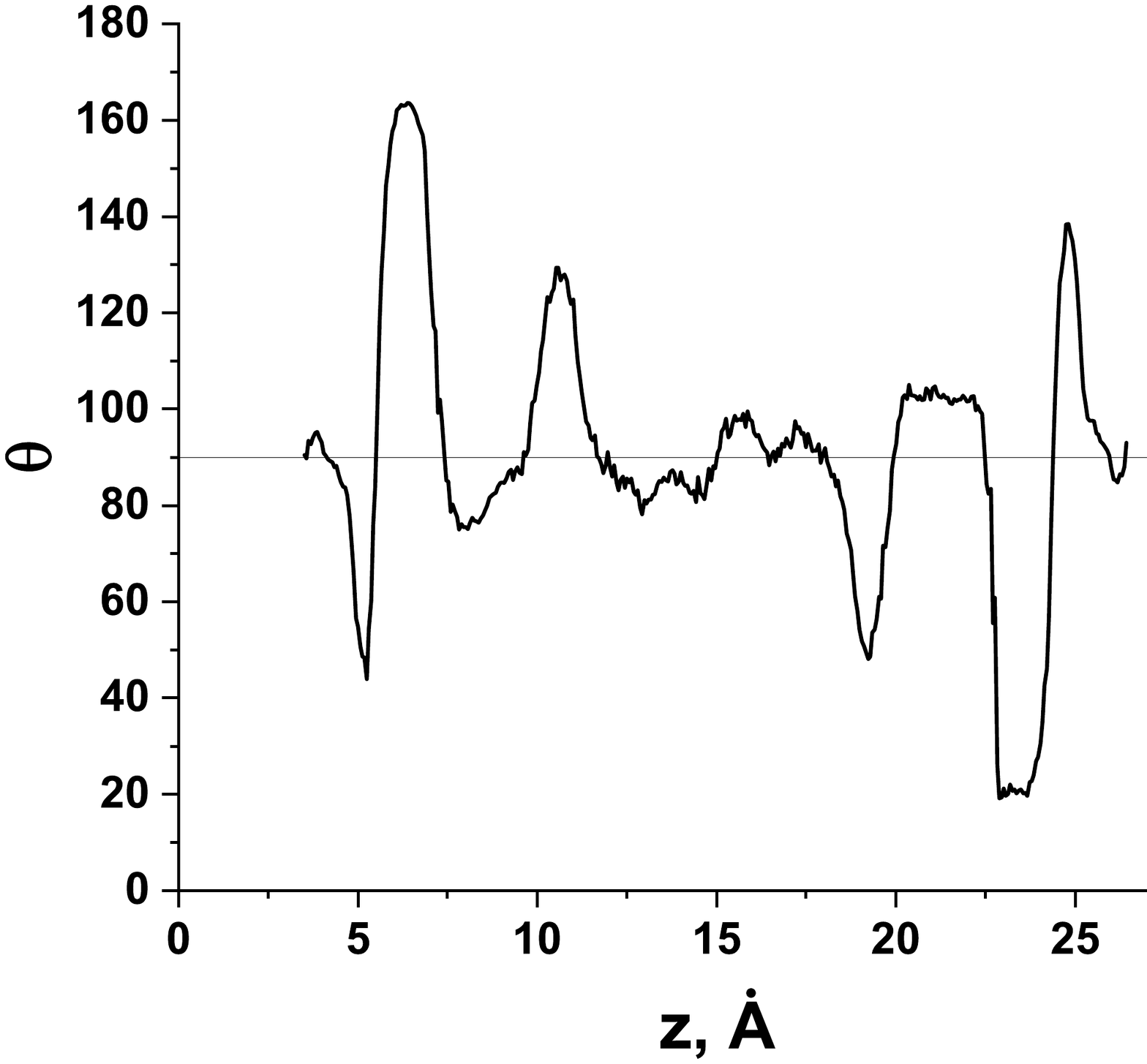}%

\caption{\label{1800} (a) A top view of the system of 1800
molecules (density $\rho=1.52$ $g/cm^3$). The atoms of graphene
are not shown for clarity. (b) A side view of the same system. (c)
The distribution of number densities of several species in the
same system. For the $F_3$ and $H_3$ species the curves are
divided by 3 in order to make them the same scale as $C_1$ and
$C_4$. (d) The distribution of angle $\theta$ along the z axis.}
\end{figure}

Finally, we show the results for density $\rho=1.52$ $g/cm^3$
($N=1800$). The results are shown in Fig. \ref{1800} (a) - (d).
The most interesting result is that the density profile of the
$C_1$ and $C_4$ species is modulated in the whole volume of the
pore, i.e. there is not any "bulk" region of the system which is
not affected by the walls. Therefore, one can claim that this
region corresponds to strong confinement. Strong modulation is
also observed in the angle distribution (panel (d)).

\begin{figure}

\includegraphics[width=6cm, height=6cm]{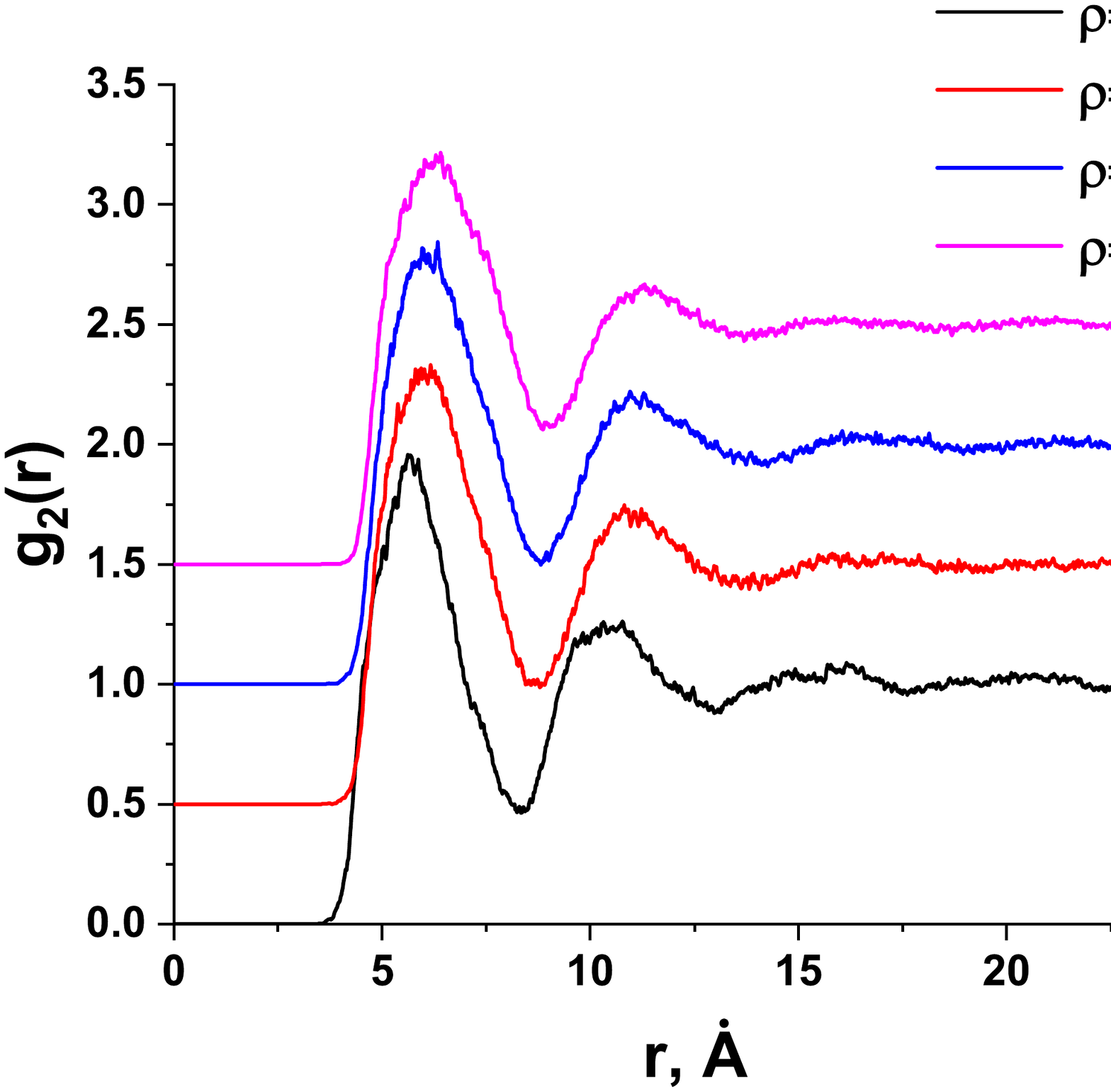}%

\caption{\label{g2} The two dimensional RDFs of the centers of
mass of the molecules in the layers next to the walls at different
densities. We shift the otherwise overlapping $g_2(r)$ curves
vertically with steps of $0.5$ for sake of clarity.}
\end{figure}

Although the density of the system increases and the system
exhibits strongly modulated density profiles, we do not observe
formation of a crystal in the system. Even in the layers next to
the walls there is no crystalline order. In order to prove it we
consider the 2d RDFs of the centers of mass of the molecules in
the layers next to the walls. The results are shown in Fig.
\ref{g2}. One can see that the first peak is almost independent of
the density. At the same time the second peak of the RDF becomes
slightly higher and wider with density and the third peak appears
at $\rho=1.52$ $g/cm^3$.

The most well studied confined liquid is definitely water
\cite{rice,phob1,phob2}. It was shown that in the case of
hydrophobic confinement water crystallized, while in the case of
hydrophilic one it experienced glass transition
\cite{phil1,phil2,phil3}. In the present case of PFB in a graphene
pore the walls appear to be philic to the liquid. As a result, we
also observe amorphization of the system rather than
crystallization.

The work was carried out using the computing resources of the
federal collective usage center "Complex for simulation and data
processing for mega-science facilities" at NRC "Kurchatov
Institute", http://ckp.nrcki.ru, and supercomputers at the Joint
Supercomputer Center of the Russian Academy of Sciences (JSCC
RAS).  The article was supported by the Ministry of Science and
Higher Education of the Russian Federation (agreement no.
075-02-2022-872).

\end{document}